\documentclass[aps,pra,twocolumn,showpacs,superscriptaddress]{revtex4}  
\usepackage{graphicx}  
\usepackage{dcolumn}   
\usepackage{bm}        
\usepackage{amssymb}   
\usepackage{clrscode}
\usepackage{mathtools}
\usepackage{multirow}
\newcommand{\CMLD}{{\sf CMLD}}
\newcommand{\QMLD}{{\sf QMLD}}
\newcommand{\DQMLD}{{\sf DQMLD}}
\newcommand{\NP}{{\sf NP}}
\newcommand{\QMA}{{\sf QMA}}

\def\cC{{\cal C}}
        \def\cE{{\cal E}}

\def\cG{{\cal G}}

\def\cL{{\cal L}}

\def\cS{{\cal S}}        \def\cT{{\cal T}}

        \def\cZ{{\cal Z}}

\def\0{{\mathbf{0}}}
\def\1{{\mathbf{1}}}
\def\2{{\mathbf{2}}}
\def\3{{\mathbf{3}}}
\def\4{{\mathbf{4}}}
\def\5{{\mathbf{5}}}
\def\6{{\mathbf{6}}}
\def\7{{\mathbf{7}}}
\def\8{{\mathbf{8}}}
\def\9{{\mathbf{9}}}

\def\bbC{\mathbb{C}}

\def\bbZ{\mathbb{Z}}

\begin{document}



\title{NP-hardness of decoding quantum error-correction codes}

\author{Min-Hsiu Hsieh}\email{minhsiuh@gmail.com}
\affiliation{Statistical Laboratory, University of Cambridge}\altaffiliation{Previous address: ERATO-SORST Quantum Computation and
Information Project, Japan Science and Technology Agency}
\author{Fran{\c c}ois Le Gall} \email{legall@is.s.u-tokyo.ac.jp}\affiliation{Department of Computer Science,
The University of Tokyo.}


\date{\today}

\begin{abstract}
Though the theory of quantum error correction is intimately related to the
classical coding theory, in particular, one can construct quantum error
correction codes (QECCs) from classical codes with the dual containing
property, this does not necessarily imply that the computational complexity of
decoding QECCs is the same as their classical counterparts. Instead, decoding QECCs
can be very much different from decoding classical codes due to the degeneracy
property. Intuitively, one expects degeneracy would simplify the decoding since
two different errors might not and need not be distinguished in order to
correct them. However, we show that general quantum decoding problem is NP-hard
regardless of the quantum codes being degenerate or non-degenerate. This
finding implies that no considerably fast decoding algorithm exists for the
general quantum decoding problems, and suggests the existence of a quantum
cryptosystem based on the hardness of decoding QECCs.
\end{abstract}

 \maketitle

\section{Introduction}\label{sec:level1}
The invention of quantum error correction codes (QECCs)
\cite{Shor95,CRSS98,Gottesman97} was one of the driving forces that boosted the
fast-growing field of quantum information and computation. Great similarities
shared between QECCs and classical codes were quickly discovered ever since,
and the later contributed significantly to the development of the former.
Particularly, a large portion of QECCs known so far is inspired or constructed
directly from classical codes. These similarities then led to a common
consensus that the general quantum decoding problem belongs to the same
computational complexity class as its classical counterpart.

In complexity theory, computational intractability can be rigorously characterized by
the concept of $\NP$-\textit{hardness}.
Formally, the complexity class $\NP$ is defined as the class of all decision
problems that can be solved by a
nondeterministic Turing machine in a number of steps polynomial in the input
length. 
A computational problem $P_1$ is said to be $\NP$-hard if it is as hard as the
hardest problems in the class $\NP$ in the following sense: the existence of a
polynomial-time algorithm for $P_1$ implies the existence of a polynomial-time
algorithm for all problems in $\NP$. A computation problem is said to be
$\NP$-complete if it is $\NP$-hard and in $\NP$.
The class of $\NP$-hard problems includes a multitude of computational tasks
believed to be intractable, such as many optimization or combinatorial problems.

It was Berlekamp \emph{et al.}~\cite{BMT78} who first showed that general
decoding problem for classical linear codes 
is $\NP$-hard. This result assured that it is unlikely that a substantially fast
algorithm for decoding the most likely classical error would exist. 
Bruck and Naor~\cite{BN90} and Lobstein~\cite{Lobstein90} then
showed that the problem remains hard even if the code is known in advance,
while Arora \emph{et al.}~\cite{ABSS97} showed that approximating the
minimal-weight solution is also $\NP$-hard.

The similarity between QECCs and classical codes breaks down regarding the
differences between how a classical error acts on a codeword and how a quantum
error acts on a code space. A striking feature of QECCs is that they can
sometimes be used to correct more errors than they can uniquely identify
\cite{ShorSmolin96}.
The feature of degeneracy in QECCs thus calls for a completely different
strategy for decoding quantum errors \cite{PoulinChung08}, and gives us hope
that there are chances that general quantum decoding can be performed
efficiently. However, in this paper, we will show that the problem of decoding
the most probable quantum error is $\NP$-hard regardless of QECCs being
degenerate or non-degenerate.

The classical McEliece cryptosystem \cite{McEliece} is considered as one of the best
candidates for post-quantum public-key cryptosystem. Its security, which is based on the hardness of decoding
general classical linear codes, has been shown recently to be also robust against
quantum Fourier sampling attacks \cite{DMR10}. Our result, showing that it is unlikely
that a substantially fast algorithm for decoding the most probable quantum error would exist,
may become a foundation of a quantum analogue of the classical McEliece cryptosystem.
Notice that a proposal of such a quantum McEliece cryptosystem has been proposed recently \cite{Fujita}.

This paper is organized as follows. In Sec.~\ref{Sec_II}, we first introduce the stabilizer formalism of quantum error correction codes, and its optimal decoding strategy. Then we relate the stabilizer formalism to the classical symplectic codes. We establish our main result -- decoding general QECCs is $\NP$-hard in Sec.~\ref{Sec_III}. We conclude the paper in Sec.~\ref{SEC_IV}.

\section{Quantum Error Correction Codes}\label{Sec_II}
Denote the set of Pauli matrices by
$\Pi\equiv\{I,X,Y,Z\}$, and define the following $n$-fold Pauli group $\cG_n$:
\[
\cG_n=\left\{[A_1\otimes A_2\otimes\cdots\otimes A_n]:A_i\in\Pi, \forall i\right\}
\]
under the multiplication operation $[A][B]=[AB]$ where $[A]=\{\alpha
A:\alpha\in\bbC, |\alpha|=1\}$ for some operator $A$. An $[[n,k]]$ QECC is a
subspace $\cC$ of size $2^k$ in the Hilbert space $\bbC^{\otimes n}$. It can be
specified as the $+1$ eigenspace of a set of commuting operators
$\{\bar{Z}_1,\cdots,\bar{Z}_{n-k}\}\subset\cG_n$ that generates under
multiplication the so-called \emph{stabilizer} group $\cS$. 

The generating set of $\cS$ can then be extended to a generating set of
$\cG_n$:
\begin{equation}\label{EQ_gent}
\cG_n=\langle \bar{Z}_i,\bar{X}_j\rangle_{1\leq i,j \leq n},
\end{equation}
such that these operators satisfy the following relations:
\begin{eqnarray}
\left[\bar{Z}_i, \bar{Z}_j\right]&=&0, \forall i,j \label{sb_basis1}\\
\left[\bar{X}_i, \bar{X}_j\right]&=&0, \forall i,j \label{sb_basis2}\\
\left[\bar{Z}_i, \bar{X}_j\right]&=&0, \forall i\neq j \label{sb_basis3}\\
\left\{\bar{Z}_i, \bar{X}_i\right\} &=& 0, \forall i. \label{sb_basis4}
\end{eqnarray}
Operators that commute with each element in the stabilizer group $\cS$ map the
code space to itself, and form a group $\cZ(\cS)$, the normalizer of $\cS$ in
$\cG_n$. Denote by $\cL\equiv \langle\bar{Z}_i,\bar{X}_j \rangle_{n-k+1\leq i,j
\leq n}$, and by $\cT\equiv\langle \bar{X}_i\rangle_{1\leq i\leq n-k}$. Errors
from $\cL$ act nontrivially on the code space $\cC$ but cannot be detected by
the error syndrome. However, errors from $\cT$ can be uniquely identified by
measuring the stabilizer $\cS$. Specifically, the $i^{th}$ element $s_i$ of the
error syndrome $\bm{s}\in(\bbZ_2)^{n-k}$ is equal to one if the error operator
$E\in\cT$ anticommutes with the $i^{th}$ generator $\bar{Z}_i$ of group $\cS$,
and is equal to zero otherwise.

The definition of \emph{degeneracy} depends on the error set $\cE$ which
the QECC is designed to correct. If two errors $E_1,E_2\in\cE$ are related by
some element in the stabilizer group $P\in\cS$, say $E_1= E_2P$, these two
errors cannot and need not be distinguished since they have the same effect on
the code space $\cC$. We then call such a QECC degenerate. On the other hand,
if each error in the error set leads to a distinct error syndrome, such a QECC is
non-degenerate. 

For any given error syndrome $\bm{s}$ (corresponding to a unique operator $T\in\cT$ in terms of the set of generators (\ref{EQ_gent})), the optimal decoding strategy is therefore to find an error $E\in\cL$ such that $\sum_{S\in\cS}\Pr(SET)$ is maximum since it will minimize the overall probability of decoding error. Notice that the probability of an error $A\in\cG_n$, $\Pr(A)$, depends on the specific channel model used. If the QECC is non-degenerate, the optimal decoding strategy reduces to finding a most likely error $E\in\cL$: $\max_{E\in\cL}\Pr(ET)$. We call such decoding strategy ``Quantum Maximum Likelihood Decoding'' (QMLD) due to its similarity to the maximally likelihood decoding in the classical setting.

There is a one-to-one correspondence between an
$[[n,k]]$ stabilizer code $\cC$ and a symplectic code $C$ of size $2^{n+k}$ in
$(\bbZ_2)^{2n}$. We will mostly use the symplectic formalism in the following
since it is more convenient to work with vectors.

Denote by $\bm{\alpha}=(\bm{z}|\bm{x})\in(\bbZ_2)^{2n}$, where
$\bm{x}=(x_1,\cdots,x_n)$ and $\bm{z}=(z_1,\cdots,z_n)$ are $n$-bit strings
with $x_i,z_i\in\bbZ_2=\{0,1\}$. There is a bijection $N:\bm{\alpha}\to
N_{\bm{\alpha}}$ that maps every symplectic vector $\bm{\alpha}$ in
$(\bbZ_2)^{2n}$ to an operator $N_{\bm{\alpha}}$ in $\cG_n$:
\begin{eqnarray*}
N_{\bm{\alpha}}&\equiv&[Z^{\bm{z}}X^{\bm{x}}]=[Z^{z_1}X^{x_1}] \otimes\cdots\otimes [Z^{z_n}X^{x_n}],
\end{eqnarray*}
where we write $Z^{\bm{z}}=Z^{z_1}\otimes\cdots\otimes Z^{z_n}$ and likewise for $X^{\bm{x}}$.

For two vectors $\bm{\alpha}=(\bm{z}|\bm{x})$ and $\bm{\beta}=(\bm{z}'|\bm{x}')$, define the symplectic
product $\odot:(\bbZ_2)^{2n}\times(\bbZ_2)^{2n}\to \bbZ_2$ to be:
\[
\bm{\alpha}\odot\bm{\beta}= \bm{z}\cdot\bm{x}'+\bm{x}\cdot\bm{z}',
\]
where $\cdot$ is the regular inner product between two vectors in $(\bbZ_2)^n$,
and $+$ is a binary addition. The symplectic product between two vectors
$\bm{\alpha}$ and $\bm{\beta}$ characterizes the commutation relation between
two operators $N_{\bm{\alpha}}$ and $N_{\bm{\beta}}$:
\[
N_{\bm{\alpha}} N_{\bm{\beta}}=(-1)^{\bm{\alpha}\odot\bm{\beta}} N_{\bm{\beta}} N_{\bm{\alpha}}.
\]

Let $S=\{\bm{\alpha}_1,\cdots,\bm{\alpha}_{n-k}\}$ be a collection of $n-k$
independent symplectic vectors in $(\bbZ_2)^{2n}$ such that
$\bm{\alpha}_i\odot\bm{\alpha}_j=0$ $\forall i,j=\{1,\cdots,n-k\}$. We can
construct the set of canonical basis vectors
$\{\bm{\alpha}_i,\bm{\beta}_j\}_{1\leq i,j \leq n}$ for $(\bbZ_2)^{2n}$ such that
\cite{BDH06,BDH061}:
\begin{eqnarray}
\bm{\alpha}_i\odot\bm{\alpha}_j&=&0, \forall i,j \label{symp_basis1}\\
\bm{\beta}_i\odot\bm{\beta}_j&=&0, \forall i,j \label{symp_basis2}\\
\bm{\alpha}_i\odot\bm{\beta}_j&=&0, \forall i\neq j \label{symp_basis3}\\
\bm{\alpha}_i\odot\bm{\beta}_i&=&1, \forall i. \label{symp_basis4}
\end{eqnarray}

Let $H$ be an $(n-k)\times 2n$ matrix where the $i$-th row vector of $H$ is
$\bm{\alpha}_i$. Define the symplectic code $C=\{\bm{\omega}\in(\bbZ_2)^{2n}:
H\odot\bm{\omega}=0\}$. It is easy to verify that
$C={\rm{span}}\{\bm{\alpha}_1,\cdots,\bm{\alpha}_n,\bm{\beta}_{n-k+1},\cdots,\bm{\beta}_n\}$.
Let $C^\perp$ be the row space of $H$, i.e., $C^\perp={\rm
span}\{\bm{\alpha}_1,\cdots,\bm{\alpha}_{n-k}\}$. Let
$L={\rm{span}}\{\bm{\alpha}_{n-k+1},\cdots,\bm{\alpha}_n,\bm{\beta}_{n-k+1},\cdots,\bm{\beta}_n\}$
and $T={\rm span}\{\bm{\beta}_1,\cdots,\bm{\beta}_{n-k}\}$. We can then
identify $\cS$, $\cL$, and $\cT$ in $\cG_n$ with $C^\perp$, $L$, and $T$ in
$(\bbZ_2)^{2n}$, respectively.

Given an error syndrome $\bm{s}\in(\bbZ_2)^{n-k}$, let $D_{\bm{s}}=\{\bm{\omega}:H\odot\bm{\omega}=\bm{s}\}$. Each error vector
$\bm{\gamma}\in D_{\bm{s}}$ can be decomposed into
$\bm{\gamma}=\bm{\gamma}_1+\bm{\gamma}_2+\bm{\gamma}_3,$ where
$\bm{\gamma}_1\in C^{\perp}$, $\bm{\gamma}_2\in L$, and $\bm{\gamma}_3\in T$.
Furthermore, the symplectic vector $\bm{\gamma}_3$ is uniquely defined by the
error syndrome $\bm{s}$:
\begin{equation}\label{vec_a3}
\bm{\gamma}_3=\sum_{i=1}^{n-k} s_i \bm{\beta}_i.
\end{equation}
For any given error syndrome $\bm{s}$, and the corresponding $\bm{\gamma}_3\in T$,
the optimal decoding strategy in the symplectic formalism is then to find a vector
$\bm{\gamma}_2\in L$ such that $\sum_{\bm{\gamma}_1\in C^\perp}\Pr(\bm{\gamma}_1+\bm{\gamma}_2+\bm{\gamma}_3)$
is maximum.

\section{Main results}\label{Sec_III}
We assume that the QECC $\cC$ is used on a Pauli channel which generates the $Z$
error and the $X$ error independently with probability $p$ (therefore the $Y$ error
occurs with probability $p^2$). Such an independency assumption has been widely used
in analysis of quantum key distribution (QKD). For example, the authors in \cite{SP00}
apply CSS-type QECCs such that the bit error and the phase error can be independently corrected.

Each error operator $N_{\bm{\gamma}}\in\cG_n$ generated by many uses of the
quantum channel occurs with probability $\Pr({\bm{\gamma}})$:
\begin{equation}\label{wt}
\Pr({\bm{\gamma}})= p^{{\rm wt}(\bm{\gamma})}(1-p)^{2n-{\rm wt}(\bm{\gamma})},
\end{equation}
where we define the function ${\rm wt}(\bm{\gamma})$ of a symplectic vector
$\bm{\gamma}=(\bm{z}|\bm{x})\in(\bbZ_2)^{2n}$ to be:
\begin{equation}\label{weight}
{\rm wt}(\bm{\gamma})=|\bm{z}|+|\bm{x}|.
\end{equation}
Here, $|\bm{a}|$ denotes the Hamming weight of a binary vector $\bm{a}$ in
$(\bbZ_2)^n$. 

As discussed in Sec.~\ref{Sec_II}, given an error syndrome $\bm{s}$ representing
an element $\bm{\gamma}_3\in T$, the optimal decoding strategy for non-degenerate QECCs is to find
the most likely error, i.e., to find a vector $\bm{\gamma}\in L$ that maximizes the
quantity $\Pr(\bm{\gamma}+\bm{\gamma}_3)$. Since we assume $p<1/2$, this is
equivalent in our setting to finding a vector $\bm{\gamma}\in L$ that minimizes
${\rm wt}(\bm{\gamma}+\bm{\gamma}_3)$, so we define the associated computational
problem as follows.\vspace{2mm}

\noindent{\sf Quantum Maximum Likelihood Decoding} (\QMLD)\vspace{1mm}

\noindent{\bf Instance}: A basis $\{\bm{\alpha}_i,\bm{\beta}_{j}\}$ of $(\bbZ_2)^{2n}$ satisfying (\ref{symp_basis1})-(\ref{symp_basis4}) \\
\hspace{17mm} and a vector $\bm{\gamma}_3\in T$.\\
\noindent{\bf Output}: \hspace{2mm}A vector $\bm{\gamma}\in L$ that minimizes ${\rm wt}(\bm{\gamma}+\bm{\gamma}_3).$\vspace{3mm}

\noindent
This decoding strategy is optimal if the
QECC is non-degenerate.
However, as mentioned in \cite{PoulinChung08}, it is not optimal
if the QECC is degenerate.
Notably, general quantum decoding deals with the error set $\cE=\cG_n$ that contains all possible errors, e.g., as
resulted by the channel model considered here in this paper. In such case, the QECC is necessarily degenerate and the optimal quantum
decoding in this case is to find the most likely set of errors that can be
corrected by the same correction operator. Given an error syndrome $\bm{s}$ representing an element $\bm{\gamma}_3\in T$,
the optimal decoding strategy is then equivalent to finding the most likely
coset $\bm{\gamma}_2+\bm{\gamma}_3+C^\perp$ in $D_{\bm{s}}$, since the operator
$N_{\bm{\gamma}_2+\bm{\gamma}_3}$ can be used to correct every error
$N_{\bm{\omega}}$, $\forall \bm{\omega}\in\bm{\gamma}_2+\bm{\gamma}_3+C^\perp$.
Let us identify  $C/C^\perp$ with the set $L$ defined above of symplectic vectors representing each coset of $C^\perp$ in $C$.
Our goal then becomes finding
$$
\arg\max_{\bm{\gamma}_2\in L}\sum_{\bm{\gamma}_1\in C^\perp}\Pr(\bm{\gamma}_1+\bm{\gamma}_2+\bm{\gamma}_3)=\hspace{20mm}
$$\vspace{-5mm}
\begin{equation}\label{optimal1}
\arg\max_{\bm{\gamma}_2\in L} \!\!\sum_{\bm{\omega}\in \bm{\gamma}_2+\bm{\gamma}_3+C^\perp}\!\!p^{{\rm wt}(\bm{\omega})}(1\!-\!p)^{2n-{\rm wt}(\bm{\omega})}.
\end{equation}
Notice that the coset containing the most likely error may not be the most likely coset determined by
Equation (\ref{optimal1}).
The associated computational problem is as follows.\vspace{2mm}

\noindent{\sf Degenerate QMLD} (\DQMLD)\vspace{1mm}

\noindent{\bf Instance}: A basis $\{\bm{\alpha}_i,\bm{\beta}_{j}\}$ of $(\bbZ_2)^{2n}$ satisfying (\ref{symp_basis1})-(\ref{symp_basis4}) \\
\hspace{17mm}\ \ \ \ \ \ \  and a vector $\bm{\gamma}_3\in T$.\\
\noindent{\bf Output}: \hspace{2mm}A symplectic vector $\bm{\gamma}\in L$ that maxi-\\
\hspace{17mm} mizes $\sum_{\bm{\omega}\in \bm{\gamma}+\bm{\gamma}_3+C^\perp}\left(p^{{\rm wt}(\bm{\omega})}(1-p)^{2n-{\rm wt}(\bm{\omega})}\right)$.\vspace{3mm}

\noindent

We say that an algorithm solves the computational problem $\QMLD$ or $\DQMLD$
in polynomial time if its running time is polynomial in $n$.
The main result of this paper is the following theorem.\vspace{2mm}
\newline
{\bf Main Theorem} \textit{
The problems $\QMLD$ and $\DQMLD$ are both $\NP$-hard.}\vspace{2mm}

\noindent 
Our result formally proves that the existence of a polynomial time algorithm for optimal quantum decoding is extremely unlikely
even in the degenerate case.

Before giving a proof of our main theorem, we review the $\NP$-completeness of classical decoding.
In the classical maximal likelihood decoding scenario, it is intuitively necessary for the receiver to search through the entire set
of $2^k$ solutions to $H \bm{n}=\bm{s}$ in order to find a solution with
minimal weight. Berlekamp \emph{et al.}~formalized this intuition and showed
that the following associated decision problem is $\NP$-complete~\cite{BMT78}.\vspace{2mm}

\noindent {\sf Classical Maximum-Likelihood Decoding} (\CMLD)\\
\noindent{\bf Instance}: An $(n-k)\times n$ matrix $A$ over $\bbZ_2$, a target\\
\hspace{16mm} vector $\bm{y}\in(\bbZ_2)^{n-k}$ and an integer $m>0$.\\
\noindent{\bf Question}: Is there a vector $\bm{w}\in(\bbZ_2)^{n}$ with $|\bm{w}|\le m$\\
\hspace{17mm} such that $A\bm{w}=\bm{y}$?\vspace{2mm}

Berlekamp \emph{et al.}~also showed that $\CMLD$ remains $\NP$-complete if $A$ is assumed
to have full row-rank (i.e., $A$ is a parity-check matrix).
It is easy to see that $\CMLD$ also remains $\NP$-complete even if $A$ is assumed to be in standard form, i.e.,
of the form
\begin{eqnarray}\label{eq:standardform}
A&=&\big[I_{n-k}\:\: P  \big]
\end{eqnarray}
for some matrix $P$ of size $(n-k)\times k$.
This is due to the fact that any linear code is permutation equivalent to a code which has a parity-check matrix in standard form, and
to the fact that this transformation can be done in polynomial time and does not change the weight distribution of the code (see for
example \cite{HP03} for a proof).

\textit{Proof of the main theorem.}---
The standard way of proving
the $\NP$-hardness of a problem $P_1$ is to prove a \textit{polynomial-time
reduction} from a $\NP$-hard problem $P_2$ to the original problem $P_1$, i.e.,
to show that any polynomial-time algorithm for $P_1$ can be used to solve in
polynomial time the problem $P_2$.

Our strategy here is to show two
polynomial-time reductions from the $\NP$-complete problem $\CMLD$: one from
$\CMLD$ to $\QMLD$ and one from $\CMLD$ to $\DQMLD$. Let $(A,\bm{y},m)$ be any
instance of the problem $\CMLD$, where $A$ is an $(n-k)\times n$ matrix over
$\bbZ_2$ of the form (\ref{eq:standardform}), $\bm{y}=(y_1,\ldots,y_{n-k})$ is
a vector in $(\bbZ_2)^{n-k}$, and $m$ is a positive integer. For convenience we
denote by $C_1\subseteq (\bbZ_2)^{n}$ the $[n,k]$ code with parity check matrix
$A$. 
We first show how to construct in polynomial time an instance
$(\{\bm{\alpha}_i,\bm{\beta}_{j}\},\bm{\gamma}_3)$ of both problems $\QMLD$ and
$\DQMLD$ embedding the information of $(A,\bm{y},m)$, and then show how a
solution to either $\QMLD$ or $\DQMLD$ can be used to compute in polynomial
time the solution to the problem~$\CMLD$.

Let us define the vector $\bm{z}=(y_1,\ldots,y_{n-k},0,\ldots,0)\in (\bbZ_2)^{n}$ and fix
$\bm{\gamma}_3=(\bm{0}|\bm{z})\in (\bbZ_2)^{2n}$.
We then define two families $\{\bm{\alpha}_i\}_{1\le i\le n}$ and $\{\bm{\beta}_i\}_{1\le i\le n}$ of vectors
in $(\bbZ_2)^{2n}$ as follows. For each $i\in\{1,\ldots,n\}$ the vector $\bm{\alpha}_i$ is the
$i$-th row of the matrix
$$
\begin{array}{ccccccll}
&\xleftrightarrow {n-k}&&\xleftrightarrow{k}&&\xleftrightarrow{n}&&\\
\multicolumn{1}{c}{\multirow{2}{*}{\Big[}}& I_{n-k}&& P&|& 0&\multicolumn{1}{c}{\multirow{2}{*}{\Big].}}&\updownarrow n-k\\
& 0&& I_k&|& 0&&\updownarrow k
\end{array}
$$
For each $i\in\{1,\ldots,n\}$ the vector $\bm{\beta}_i$ is the
$i$-th row of the matrix
$$
\begin{array}{ccccccll}
&\xleftrightarrow {n}&&\xleftrightarrow{n-k}&&\xleftrightarrow{k}&&\\
\multicolumn{1}{c}{\multirow{2}{*}{\Big[}}& 0&|& I_{n-k}&& 0&\multicolumn{1}{c}{\multirow{2}{*}{\Big].}}&\updownarrow n-k\\
& 0&|& P^T&& I_k&&\updownarrow k
\end{array}
$$
Notice that $\{\bm{\alpha}_i\}_{1\le i\le n}$ and $\{\bm{\beta}_i\}_{1\le i\le n}$ satisfy (\ref{symp_basis1})-(\ref{symp_basis4}).
The reason why $\bm{\alpha}_i\odot\bm{\beta}_{n-k+\ell}=0$ for $i\in\{1,\ldots,n-k\}$
and $\ell\in\{1,\ldots,k\}$ may be unclear. This is because
\[
\bm{\alpha}_i\odot\bm{\beta}_{n-k+\ell}=[P^T]_{\ell i}+[P]_{i\ell}=0
\]
since we are working over the binary field. The families $\{\bm{\alpha}_i\}_{1\le i\le n}$ and $\{\bm{\beta}_i\}_{1\le i\le n}$ define the following
subsets of~$(\bbZ_2)^{2n}$:
\begin{equation}\label{eq:C_perp}
C^\perp=\left\{(\bm{u}|\0): \forall \bm{u}\in C_1^\perp\right\},
\end{equation}\vspace{-7mm}
\begin{equation}\label{eq:C}
C=\left\{
(\bm{u}|\bm{v}): \forall \bm{u}\in(\bbZ_2)^{n}, \text{and}\ \forall \bm{v}\in C_1\right\},
\end{equation}\vspace{-7mm}
\begin{equation}\label{eq:T}
T\!=\!\{(\bm{0}|b_1,\ldots,b_{n-k},0,\ldots,0):b_1,\ldots,b_{n-k}\in\bbZ_2\}.
\end{equation}
Observe that $\bm{\gamma}_3\in T$ and
notice that
\begin{equation}\label{eq:C_over_D}
\bm{\gamma}_3+L\!=\left\{(\bm{u}|\bm{v}): \forall \bm{u}\in R, \text{and}\ \forall \bm{v}\in \bm{z}+C_1\right\},
\end{equation}
where $R$ is the subset of $(\bbZ_2)^n$ defined as $R=\{(0,\ldots,0,c_{1},\ldots,c_{k}):c_{1},\ldots,c_{k}\in\bbZ_2\}$.
Moreover, for any $\bm{\omega}\in \bm{\gamma}_3+L$
written as $\bm{\omega}=(\bm{u}|\bm{v})$ with $\bm{u},\bm{v}\in(\bbZ_2)^n$, we have
\begin{eqnarray}\label{tmp1}
{\rm wt}(\bm{\omega}+\bm{\mu}) &=& |\bm{v}|+{\rm wt}(\bm{\mu}+(\bm{u}|\0))
\end{eqnarray}
for all $\bm{\mu}\in C^\perp$.

The second step of the proof is specific to the problem considered.
Let us first consider the reduction from $\CMLD$ to $\QMLD$, which is the simplest case.
If we run an algorithm for $\QMLD$ on the instance $(\{\bm{\alpha}_i,\bm{\beta}_{j}\},\bm{\gamma}_3)$ just constructed, the output will be
\begin{eqnarray*}
\hat{\bm{\gamma}}=\arg\min_{\bm{\gamma}\in L} \left[{\rm wt}(\bm{\gamma}+\bm{\gamma}_3)\right].
\end{eqnarray*}
Let us write $\hat{\bm{\gamma}}=(\hat{\bm{u}}|\hat{\bm{v}})$ and notice that necessarily $\hat{\bm{u}}=\bm{0}$.
Then, using the fact that $A$ is the parity check matrix of $C_1$ and $A\bm{z}=\bm{y}$, we obtain:
\begin{eqnarray*}
{\rm wt}(\hat{\bm{\gamma}}+\bm{\gamma}_3)&=&|\hat{\bm{v}}+\bm{z}|\\
&=&\textrm{min}_{\bm{v}\in \bm{z}+C_1}[\: |\bm{v}|\:]\\
&=&\textrm{min}_{\substack{\bm{v}\in (\bbZ_2)^n\\\textrm{s.~t. } A\bm{v}=\bm{y}}}[\: |\bm{v}|\:].
\end{eqnarray*}

Let us now consider the reduction from $\CMLD$ to $\DQMLD$.
If we run an algorithm for $\DQMLD$ on the instance $(\{\bm{\alpha}_i,\bm{\beta}_{j}\},\bm{\gamma}_3)$, the output will be
\begin{eqnarray*}
\hat{\bm{\gamma}}=\arg\max_{\bm{\gamma}\in L} \left[\sum_{\bm{\omega}\in \bm{\gamma}+\bm{\gamma}_3+C^\perp} p^{{\rm wt}(\bm{\omega})}(1-p)^{2n-{\rm wt}(\bm{\omega})}\right].
\end{eqnarray*}
Let us write $\hat{\bm{\gamma}}=(\hat{\bm{u}}|\hat{\bm{v}})$ with $\hat{\bm{u}},\hat{\bm{v}}\in(\bbZ_2)^n$.
Observe that, for any $\bm{\tau}=(\bm{u}|\bm{v})\in  \bm{\gamma}_3+L$, Equality~(\ref{tmp1}) implies that
\begin{eqnarray}\label{expr1}
\left[\sum_{\bm{\mu}\in C^\perp} p^{{\rm wt}(\bm{\mu}+\bm{\tau})}(1-p)^{2n-{\rm wt}(\bm{\mu}+\bm{\tau})}\right]
=\kappa_{\bm{u}}\cdot \lambda_{\bm{v}},
\end{eqnarray}
where
\begin{eqnarray*}
\kappa_{\bm{u}}&=&\sum_{\bm{\mu}\in C^\perp} p^{{\rm wt}(\bm{\mu}+(\bm{u}|\0))}(1-p)^{2n-{\rm wt}(\bm{\mu}+(\bm{u}|\0))},\textrm{ and}\\
\lambda_{\bm{v}}&=&\left(\frac{p}{1-p}\right)^{|\bm{v}|}.
\end{eqnarray*}
Notice that Expression~(\ref{expr1}) reaches its maximum over $\bm{\gamma}_3+L$
for the value $\bm{\tau}=\hat{\bm{\gamma}}+ \bm{\gamma}_3=(\hat{\bm{u}}|\hat{\bm{v}}+\bm{z})$.
Due to properties of the set $\bm{\gamma}_3+L$ 
immediate from Equation (\ref{eq:C_over_D}),
it is easy to see
that the term $\lambda_{\bm{v}}$ also reaches its maximum for the value $\hat{\bm{\gamma}}+ \bm{\gamma}_3$, i.e.,
\begin{eqnarray*}
\lambda_{\hat{\bm{v}}+\bm{z}}=\max_{(\bm{u}|\bm{v})\in \bm{\gamma}_3+L}  \left[ \lambda_{\bm{v}}\right].
\end{eqnarray*}
Since the term $\lambda_{\bm{v}}$ is maximized for a vector $\bm{v}$ of minimal weight (because $p<1/2$),
we conclude that
 \begin{eqnarray*}
|\hat{\bm{v}}+\bm{z}|&=&\textrm{min}_{(\bm{u}|\bm{v})\in \bm{\gamma}_3+L}[\: |\bm{v}|\: ]\\
&=&\textrm{min}_{\bm{v}\in \bm{z}+C_1}[\: |\bm{v}|\:]\\
&=&\textrm{min}_{\substack{\bm{v}\in (\bbZ_2)^n\\\textrm{s.~t. } A\bm{v}=\bm{y}}}[\: |\bm{v}|\:],
\end{eqnarray*}
where the second equality comes from Equation (\ref{eq:C_over_D}).

Then, for both $\QMLD$ and $\DQMLD$, the obtained value~$\hat{\bm{v}}$ can be
used to solve the original instance of $\CMLD$ in a straightforward way: there
exists a vector $\bm{w}\in(\bbZ_2)^{n}$ with $|\bm{w}|\le m$ such that
$A\bm{w}=\bm{y}$ if and only if $|\hat{\bm{v}}+\bm{z}|\le m$. To summarize, if
there exists a polynomial-time algorithm solving either $\QMLD$ or $\DQMLD$, it
will of course work in polynomial-time for the instance constructed above, and
then solves in polynomial-time the problem $\CMLD$. This shows that the
problems $\QMLD$ and $\DQMLD$ are $\NP$-hard, and completes the proof of our
main theorem.

Though the independency assumption of the bit errors and phase errors in our channel model
leads to a great simplification (because all the probabilities involving the degeneracy of the code can be factored out), we shall stress that
the resulting decoding problem still captures the quantumness in the sense that the degeneracy is preserved.
Equivalently, such problem can be viewed as classical coset decoding, where the goal is to find a coset leader of a particular classical code, instead of simply classical ML decoding.

\section{Conclusion}\label{SEC_IV}
In this paper, we rigorously formulated and proved that general quantum decoding is NP-hard. This settles the longstanding problem of classifying the computational problem of the general quantum decoding regardless of the QECCs being degenerate or non-degenerate. Our result also implies that classically finding a target coset representative is hard since QECCs are instances of classical coset codes.
%
Finally, our result established the theoretic foundation of the development of a quantum McEliece cryptosystem.

One interesting follow-up work is to investigate hardness with respect to the complexity class defined in terms of a model of quantum computation since quantum decoding problems are genuine quantum information processing tasks. A first target may be the quantum complexity class $\QMA$ (see, e.g., \cite{Watrous00}), which is often considered as a natural quantum version of $\NP$.
One can indeed ask if the decoding problems considered in these papers are $\QMA$-hard as well. Finally, even if the problems $\QMLD$ and $\DQMLD$ are $\NP$-hard, as shown in this paper, it would be
desirable to develop algorithms for them: algorithms with subexponential time complexity, approximation algorithms, or algorithms working for special cases.

\section{Acknowledgement}
The authors would like to thank Todd Brun and Mark Wilde for
useful discussion. M.H. acknowledges support from the European Community's Seventh
Framework Programme (FP7/2007-2013) under grant agreement number 213681.
F.L.G. acknowledges support from the Japan Society for the Promotion of Science,
under the grant-in-aid for research activity start-up number 22800006.


\begin{thebibliography}{99}
\bibitem{Shor95} P. W. Shor. Phys. Rev. A {\bf 52}, (1995).

\bibitem{CRSS98} A. R. Calderbank, E. M. Rains, P. W. Shor, and N. J. A. Sloane. IEEE Trans. Inf. Theory {\bf 44}, (1998).

\bibitem{Gottesman97} D. Gottesman, PhD thesis, California Institute of Technology, (1997).

\bibitem{BMT78}
E. R. Berlekamp, R. J. McEliece, and H. C. A. van Tilborg.  IEEE Trans. Inf. Theory {\bf 24}, (1978).

\bibitem{BN90}
J. Bruck and M. Naor.  IEEE Trans. Inf. Theory {\bf 36}, (1990).

\bibitem{Lobstein90}
A.C. Lobstein. IEEE Trans. Inf. Theory {\bf 36}, (1990).

\bibitem{ABSS97}
S. Arora, L. Babai, J. Stern, and Z. Sweedyk. J. Computer System Sci. {\bf  54}, (1997).

\bibitem{ShorSmolin96} P. Shor and J. Smolin. quant-ph/9604006.


\bibitem{PoulinChung08} D. Poulin and Y.-J. Chung. QIC {\bf 8}, (2008).

\bibitem{McEliece} R. J. McEliece. JPL DSN Progress Report, 42--44, (1978).

\bibitem{DMR10} H. Dinh, C.~Moore, A.~Russell. arXiv:1008.2390.

\bibitem{Fujita} H. Fujita. In preparation.

\bibitem{BDH06} T.A. Brun, I. Devetak, and M.-H. Hsieh. \textit{ New Trends
    in Mathematical Physics Selected contributions of the XVth International
    Congress on Mathematical Physics}, (2009).
\bibitem{BDH061}T. Brun, I. Devetak, and M.-H. Hsieh. \emph{Science} {\bf 314}, (2006).

\bibitem{SP00} P. W. Shor and J. Preskill. Phys. Rev. Lett. {\bf{85}}, (2000).

\bibitem{HP03}
W.C. Huffman and V. Pless. \textit{Fundamentals of Error Correcting Codes}, (2003).

\bibitem{Watrous00}
J. Watrous. \textit{Proceedings of the 41st Annual IEEE Symposium on Foundations of Computer Science}, (2000).

\end{thebibliography}
\end{document}